\documentstyle[12pt]{article}
\input epsf
\def\wlog#1{} 
\catcode`\@=11

\def\({\relax\ifmmode[\else$[$\nobreak\hskip.3em\fi}
\def\){\relax\ifmmode]\else\nobreak\hskip.2em$]$\fi}

\def\gappr{\mathpalette\under@rel{>\approx}}
\def\lappr{\mathpalette\under@rel{<\approx}}
\def\gsim{\mathpalette\under@rel{>\sim}}
\def\lsim{\mathpalette\under@rel{<\sim}}
\def\under@rel#1#2{\under@@rel#1#2}

\def\under@@rel#1#2#3{\mathrel{\mathop{#1#2}\limits_{#1#3}}}

\def\under@@rel#1#2#3{\mathrel{\vcenter{\hbox{$%
  \lower3.8pt\hbox{$#1#2$}\atop{\raise1.8pt\hbox{$#1#3$}}%
  $}}}}

\def\parenbar{\mathpalette\p@renb@r}
\def\p@renb@r#1#2{\vbox{%
  \ifx#1\scriptscriptstyle \dimen@.7em\dimen@ii.2em\else
  \ifx#1\scriptstyle \dimen@.8em\dimen@ii.25em\else
  \dimen@1em\dimen@ii.4em\fi\fi \offinterlineskip
  \ialign{\hfill##\hfill\cr
    \vbox{\hrule width\dimen@ii}\cr
    \noalign{\vskip-.3ex}%
    \hbox to\dimen@{$\mathchar300\hfil\mathchar301$}\cr
    \noalign{\vskip-.3ex}%
    $#1#2$\cr}}}

\def\veq{\afterassignment\v@eq \dimen@}
\def\v@eq{$$\vcenter to\dimen@{}$$}

\def\veqn{\afterassignment\v@eqn \dimen@}
\def\v@eqn{$$\vcenter to\dimen@{}\eqn$$}

\def\heq{\afterassignment\h@eq \dimen@}
\def\h@eq{$\hbox to\dimen@{}$ }

\def\wlog{\immediate\write\m@ne} 
\catcode`\@=12 

\def\zerslash{\hbox{0\kern-.5em/}}
\def\lappr{\lsim}
\def\gappr{\gsim}
\def\sqr#1#2#3{{\vcenter{\hrule height.#3ex\hbox{\vrule width.#2ex height#1ex
    \kern#1ex\vrule width.#3ex}\hrule height.#2ex}}}

\def\angleto{\vrule width.035em height2.1ex depth-.56ex\unskip\kern-.6ex\to}
\mathchardef\perslsh=47
\def\perchc#1{{\raise.4ex\hbox{$\mkern4mu#1{\it\perslsh}_
             {\mkern-5mu\scriptscriptstyle{{\rm o}\!{\rm o}}}^
             {\mkern-12.8mu\scriptscriptstyle{\rm o}}$}}}

\newbox\struttbox
\setbox\struttbox=\hbox{\vrule height1.65ex depth.485ex width0pt}
\def\strutt{\relax\ifmmode\copy\struttbox\else\unhcopy\struttbox\fi}
\def\stru#1#2{\relax\ifmmode\hbox{\vrule height#1 depth#2 width0pt}
\else\vrule height#1 depth#2 width0pt\fi}

\mathchardef\smallleft=300
\mathchardef\smallright=301

\def\ronum#1{\uppercase\expandafter{\romannumeral#1}}
\def\ronuml#1{\expandafter{\romannumeral#1}}

\newfont{\bm}{cmmib10 scaled 1200}
\mathchardef\pls=43
\mathchardef\mns=512
\mathchardef\plm=518
\mathchardef\eql=61

\def\ev{{\rm ~e}\kern-1.pt{\rm V}}
\def\kev{{\rm ~ke}\kern-1.pt{\rm V}}
\def\mev{{\rm ~Me}\kern-1.pt{\rm V}}
\def\gev{{\rm ~Ge}\kern-1.pt{\rm V}}
\def\tev{{\rm ~Te}\kern-1.pt{\rm V}}

\def\etamax{\eta_{{\rm max}}}

\newcommand{\pom}{{I\!\!P}}

\begin{document}
\title{
{\Large \bf {Production of Large Rapidity Gap Events in $ep$
Interactions at HERA}}
\footnote{ Talk given at the VIth Blois Workshop on Frontiers in Strong
Interactions, Blois, France, June 20--24, 1995}\\
}

\author{Halina Abramowicz\\
School of Physics, Tel--Aviv University\\[0.4cm]
Representing the ZEUS Collaboration\\
}
\date{\ }
\maketitle

\vspace{5cm}

\begin{abstract}
This is a short review of the properties of electron proton
interactions characterized by the presence of large rapidity gaps
(LRG) in the measured hadronic final state as obtained by the ZEUS
Collaboration at the HERA Collider.  In the deep inelastic neutral
current $ep$ interactions, the factorization properties of the LRG
events interpreted as due to the diffractive dissociation of the
virtual photon are compatible with expectations from the Regge
phenomenology of soft interactions. The measurement of deep inelastic
scattering combined with results from photoproduction of high $p_T$
jets are successfully interpreted in terms of a factorizable Pomeron
consisting of quarks and with a substantial contribution of a gluonic
component.  The first hints of a more complicated nature of the
Pomeron are observed in the deep inelastic exclusive $\rho^o$
production, where a strong increase of the production cross section
with energy is observed relative to the measurements of the NMC
Collaboration at lower energy.
\end{abstract}

\vspace{2cm}

\thispagestyle{empty}
\setcounter{page}{0}

\newpage

\section{Introduction}

The study of $ep$ interactions at the HERA collider, where 26.7 GeV
electrons collide with 820 GeV protons, has lead to the observation of
many interesting effects, with highlights such as the strong rise of
the proton structure function $F_2$ in the region of low Bjorken $x$
even at moderate momentum transfers $Q^2 \sim 10
\gev$~\cite{H1F2,ZEUSF2} and the production of large rapidity gap
(LRG) events in the deep inelastic neutral current
interactions~\cite{ZEUSLRG,H1LRG} and in high $p_T$ jet
photoproduction~\cite{ZEUSLRGPH,H1LRGPH}.

Particles produced in inelastic, non diffractive hadron hadron
interactions, are known to populate a cylindrical phase space (with
limited $p_T$) available for their production, with three
characteristic domains: the two fragmentation regions corresponding to
the initial beam particles and the central region. A similar picture
emerges in $ep$ interactions~\cite{HFS} when viewed from the
$\gamma^*p$ frame after removing the scattered electron.  On the other
hand particles originating from a diffractively excited $\gamma^*$
will only populate the photon fragmentation region and lead to a large
rapidity gap (LRG) extending through (part of) the central and proton
fragmentation regions.

The initial study of the LRG events observed in NC DIS interactions at
HERA~\cite{ZEUSLRG,H1LRG,ZEUSLRGJET} led to the conclusion that: (1)
the $Q^2$ dependence was consistent with originating from a leading
twist effect, (2) the ratio of LRG events to all the events was almost
constant with $W$, (3) a small fraction of about 3\% of the LRG events
had large $p_T$ jets in the final state, thus leading to a picture of
predominantly soft hadronic configurations. The emerging picture was
consistent with the assumption of a DIS scattering on a quark
originating from a Pomeron like object, in good agreement with models
such as that of Ingelman-Schlein~\cite{IS},
Donnachie--Landshoff~\cite{DLpom} or of Nikolaev-Zakharov~\cite{NZ}.

\section{Large Rapidity Gap Events in the ZEUS detector}

The ZEUS detector~\cite{ZEUSDET} consists of the vertex detector, the
central tracking chamber, the high resolution uranium/scintillator
calorimeter (CAL) and a muon spectrometer consisting in turn of an iron
backing calorimeter sandwiched between precision muon chambers. In the
forward (defined along the momentum vector of the incoming proton) and
rear (defined by the momentum vector of the incoming electron)
directions the tracking is supplemented by the forward and rear
tracking chambers.  The maximal rapidity acceptance is given by the
CAL and extends from 4.2 to -3.8 units of pseudorapidity $\eta$ ($\eta
= -\ln
\tan\frac{\theta}{2}$ with $\theta$ the polar angle measured with respect
to the proton direction).

A leading proton spectrometer, LPS, consisting of a chain of roman
pots surrounding the beam pipe is located in the forward region to tag
and measure diffractive photon dissociation reactions.  A forward
neutron calorimeter, FNC, has been designed to tag high energy
neutrons originating from the $ep$ interactions.

The luminosity is determined by measuring the integrated Bethe-Heitler
photon spectrum ($ep \rightarrow ep\gamma$), measured in the
luminosity gamma detector located downstream of the electron beam.

The $ep$ interactions are classified as NC DIS if the scattered
electron is detected in the CAL and as photoproduction otherwise. The
separation occurs typically for photon virtualities of $Q^2 =
4$~GeV$^2$.

In this presentation the operational definition of diffractive events
is adopted following Bjorken~\cite{BjLRG}: '' A diffractive process
occurs if and only if there is a large rapidity gap in the
produced particle phase space which is not exponentially suppressed.''
In the HERA regime diffractive photon dissociation can be tagged by a
presence of a LRG for masses up to about 20 GeV which can be fully
contained in the acceptance region of the central detector.
\begin{figure}[htb]
\begin{minipage}[t]{80mm}
\epsfxsize=80mm
\epsfclipon
\epsffile[40 180 520 640]{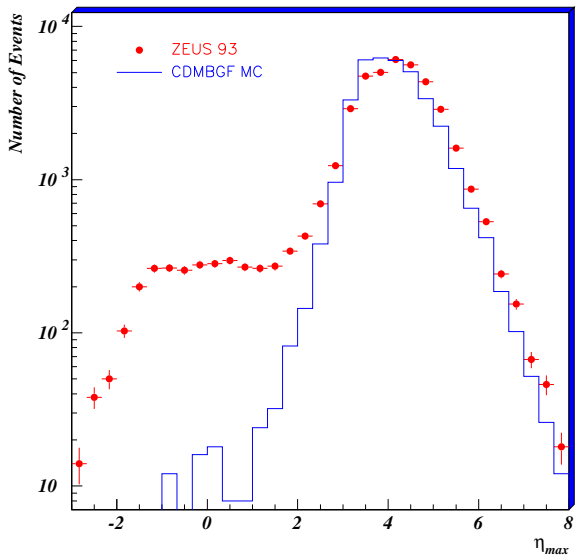}
\end{minipage} \  \
\begin{minipage}[b]{80mm}
\caption{\label{etadis}{ \sl
Distribution of $\etamax$ as defined in the text
for DIS NC events in the data (dots) and in the MC.
(histogram) }}
\vspace{20mm}
\end{minipage}
\end{figure}
To select large rapidity gap events we define for each event
$\eta_{\rm max}$ as the $\eta$ of an energy deposit in the detector
above 400 MeV closest to the proton direction. The distribution of
$\eta_{\rm max}$ for selected DIS NC events with $Q^2>8$~GeV$^2$ is
presented in figure~\ref{etadis} and compared with a DIS Monte Carlo (MC)
based on the Color Dipole Model (ARIADNE~\cite{ariadne}) for the
production of the hadronic final state. A clear excess of events with
large rapidity gaps is observed, in the region of $\eta_{\rm max}<2$
which corresponds to an effective rapidity gap of more than 2 units of
$\eta$ in the detector and possibly more than 5 units of $\eta$
relative to the initial proton. In the MC the LRG events are strongly
suppressed in this region.

\section{Diffractive Deep Inelastic Scattering}

The LRG events can be described by the following set of inclusive
variables: the negative of the four momentum transfer squared between
the incoming and the scattered electron -- $Q^2$; the fraction of the
proton momentum carried by the struck quark determined by Bjorken $x$;
the invariant mass of the hadronic final state $W$ equal to the center
of mass energy of the $\gamma^*p$ system; the square of the momentum
transfer, $t$, between the initial and final proton; the invariant
mass of the hadrons excluding the final state proton, $M_X$; and
finally the fraction of the proton momentum carried by the (generic)
pomeron $x_{\pom}$. It is customary to introduce a variable which
denotes the fraction of the pomeron momentum carried by the struck
quark, $\beta = x/x_{\pom}$.

Assuming $t \simeq 0$ and neglecting the mass of the proton
\begin{equation}
x_{\pom} = \frac{Q^2+M_X^2}{Q^2+W^2} \, , \ \ \ \
\beta = \frac{Q^2}{Q^2+M_X^2} \, . \nonumber
\end{equation}
To study the factorization properties of LRG events it is convenient to
rewrite the differential DIS $ep$ cross section as a function of $Q^2,
\beta$ and $x_{\pom}$,
\begin{equation}
\frac{d^3\sigma}{d\,\beta d\,Q^2 d\, x_{\pom} } =
\frac{2\pi \alpha^2}{\beta  Q^2} \left[ 1+(1-y)^2\right]
F^{D(3)}_2 (\beta,Q^2,x_{\pom}) \, , \nonumber
\end{equation}
where $y=\frac{Q^2}{xs}$ with $s$ the $ep$ center of mass energy
squared and $F^{D(3)}_2$ denotes the contribution of LRG events with a
given $x_{\pom}$ integrated over $t$ to the $F_2$ structure function
of the proton for a given $Q^2$ and $x=\beta x_{\pom}$. For simplicity
the contribution from the $F_L$ structure function has been omitted in
the above formula.

Should the mechanism of diffractive production in DIS be similar to
that ruling the soft hadron hadron interactions, the $F^{D(3)}_2$ is
expected to factorize into a flux factor $f_{\pom}$ and the structure
function of the pomeron $F_2^{\pom}(\beta,Q^2)$: $F^{D(3)}_2 =
f_{\pom}(x_{\pom}) \times F_2^{\pom}(\beta,Q^2)$. Moreover if the LRG
events are induced by the $\pom$ trajectory of the Regge phenomenology,
$\alpha_{\pom}(t) = 1 + \epsilon + \alpha' t$
with $\epsilon = 0.08$~\cite{DL} and $\alpha'=0.25$ GeV$^{-2}$, the
$x_{\pom}$ dependence of the flux integrated over $t$ is expected to be
\[
f_{\pom}(x_{\pom}) \sim \frac{1}{x_{\pom}^{1.16}} \cdot x_{\pom}^{0.06
\div 0.04} \, .
\]
The term of $x_\pom^{0.06 \div 0.04}$ comes from the slope of
the $\pom$ trajectory and the variation depends on the assumed
$t$ dependence of the diffractive peak ($\sim e^{Bt}$ with $B=4.5 \div
8$ GeV$^{-2}$).  The structure function $F_2^{\pom}$ would then depend
on the partonic structure of the pomeron. This is the essence of the
Ingelman--Schlein~\cite{IS} and Donnachie--Landshoff~\cite{DLpom}
models for diffraction in DIS. Note that the separation of $F^{D(3)}_2$
into a flux term and $F_2^{\pom}$ is not unique since the
normalization of $F_2^{\pom}$ is not known.

In the Nikolaev-Zakharov model~\cite{NZ} the pomeron is modelled by a
two gluon exchange with an addition of multi--gluon exchanges parameterized
by the triple Regge formula.  This leads to an explicit factorization
breaking, although the effect is small. In the model of Capella et
al.~\cite{Capella} the flux of the $\pom$ is determined by the
intercept of the bare pomeron and the structure function of the $\pom$
is related to that of the deuterium.

The results reported here correspond to an integrated luminosity of
0.54 pb$^{-1}$ collected during the 1993 running period.  The
selection of events on the basis of the $\etamax \leq 1.5$ cut leads
to a very clean sample of LRG events, yet it restricts the analysis to
relatively small masses (typically $M_X \lsim 10~
\gev$). The efficiency to select larger masses can be improved at
the expense of an increased background from the non-diffractive DIS
events. In the analysis presented here the LRG events were selected
requiring $\etamax < 2.5$ and $\cos \theta_h = \sum_i p_{zi} / \sum_i
\vec{p_i} < 0.75$, where the sum runs over all calorimeter clusters
assigned to the hadronic final state. The latter cut allows to
separate large mass diffractive events which do not exhibit a LRG in
the calorimeter from non-diffractive events in which due to the
fragmentation of the proton system the hadronic energy measured in the
forward calorimeter is large and thus $\cos \theta_h \simeq 1$.  The
remaining background from non-diffractive events is estimated with the
DIS MC based on the Color Dipole Model including boson gluon fusion
diagrams~\cite{ariadne} and subtracted. The contribution from double
diffractive dissociation is estimated to be $15 \pm 10 \%$ and is not
subtracted. It is assumed to be independent of $x_{\pom}$.  Only
events with $0.08<y<0.5$ were used in the analysis. The upper cut is
applied in order to limit the contribution from the unknown longitudinal
structure function for the LRG component.
\begin{figure}[htb]
\begin{minipage}[b]{90mm}
\epsfxsize = 90mm
\epsfclipon
\epsffile[30 200 550 650]{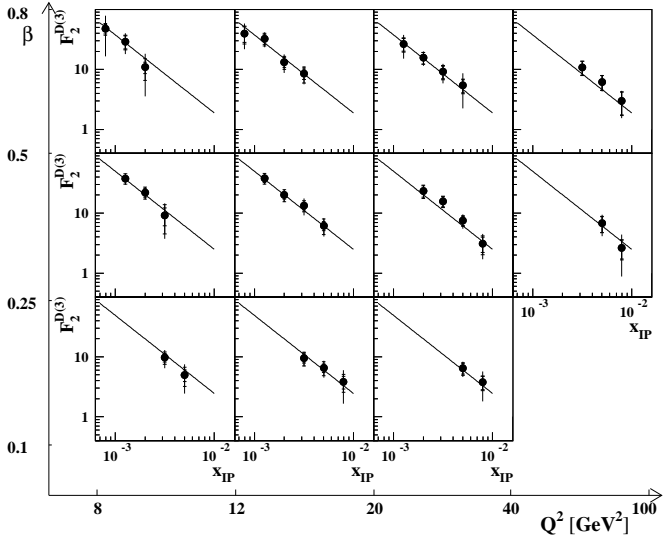}
\end{minipage} \ \
\begin{minipage}[b]{70mm}
\caption{\label{f2d3}{\protect \sl
$F_2^{D(3)}(\beta, Q^2, x_{\pom})$ as a function of $x_{\pom}$ for
fixed $\beta$ and $ Q^2$, compared to the parameterization discussed in
the text. The statistical and systematic errors are shown separately
as well as the total error.  Note that the data include an estimated
15\% contribution due to double dissociation.  The overall
normalization uncertainty of 3.5\% due to the luminosity uncertainty
is not included.  }}
\vspace{15mm}
\end{minipage}
\end{figure}
The results of the $F^{D(3)}_2$ determination are presented in bins of
$\beta$ and $Q^2$ as a function of $x_{\pom}$ in figure~\ref{f2d3}.
The data can be well represented by a fit of the form
\[
F^{D(3)}_2 (\beta,Q^2,x_{\pom}) = \left(\frac{1}{x_{\pom}}\right)^a
C_{\beta,Q^2} \, ,
\]
where $a$ is kept fixed for all $\beta$, $Q^2$ bins and $C$ is a
constant allowed to vary from bin to bin. The result of the fit yields
\[
a= 1.30 \pm 0.08 ({\rm stat.})^{+0.08}_{-0.14} ({\rm syst.})
\]
All the data are well described by
\[
F^{D(3)}_2 (\beta,Q^2,x_{\pom}) = 0.018
\left(\frac{1}{x_{\pom}}\right)^{1.30}
\left(\beta(1-\beta) + \frac{0.57}{2}(1-\beta)^2 \right) \, ,
\]
with no $Q^2$ dependence. The result of the fit is presented in
figure~\ref{f2d3}.  It is  within errors compatible  with
diffractive production being driven by the soft $\pom$.
We define $\tilde{F}_2^D (\beta,Q^2)$ as
\[
\tilde{F}_2^D (\beta,Q^2)=\int \limits^{10^{-2}}_{6.3 \cdot 10^{-4}}
F^{D(3)}_2 (\beta,Q^2,x_{\pom}) d\,x_{\pom}
\]
\begin{figure}[htb]
\begin{minipage}[b]{90mm}
\epsfxsize = 90mm
\epsfclipon
\epsffile[30 200 550 690]{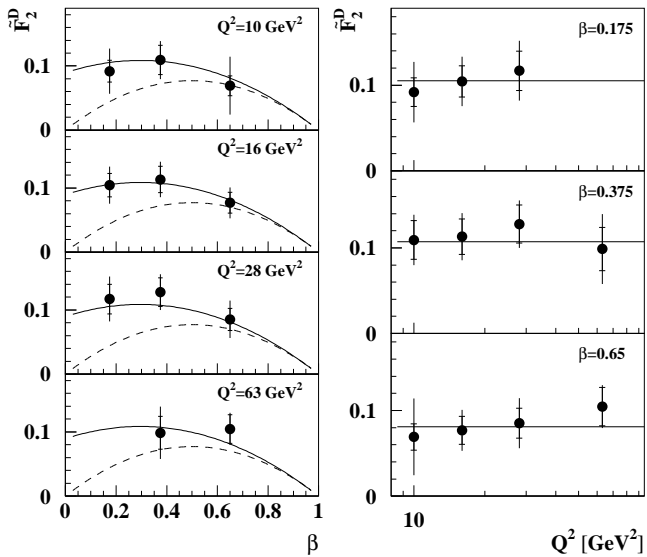}
\end{minipage} \ \
\begin{minipage}[b]{70mm}
\caption{\label{f2bq2}{\protect\sl
$\tilde{F}_2^D$($\beta, Q^2$) as a function of $\beta$ for fixed $Q^2$
and as a function of $Q^2$ for fixed $\beta$. The full line indicates
the parameterization discussed in the text. The $\beta(1-\beta)$
contribution is indicated by the dashed line.}}
\vspace{20mm}
\end{minipage}
\end{figure}
The $\beta$ dependence for fixed $Q^2$ values and the $Q^2$ dependence
for fixed $\beta$ values of $\tilde{F}_2^D (\beta,Q^2)$ is presented
in figure~\ref{f2bq2}. The $\beta$ distribution is compared to the
form $\beta(1-\beta)$ expected in the splitting of the $\pom$ into a
$q\bar{q}$ pair. A clear excess of low $\beta$ events is
observed. Within the statistical and systematic errors the result is
compatible with no $Q^2$ dependence even at the largest value of
$\beta$.
\begin{figure}[htb]
\begin{minipage}[b]{90mm}
\epsfxsize = 90mm
\epsfclipon
\epsffile[30 200 550 650]{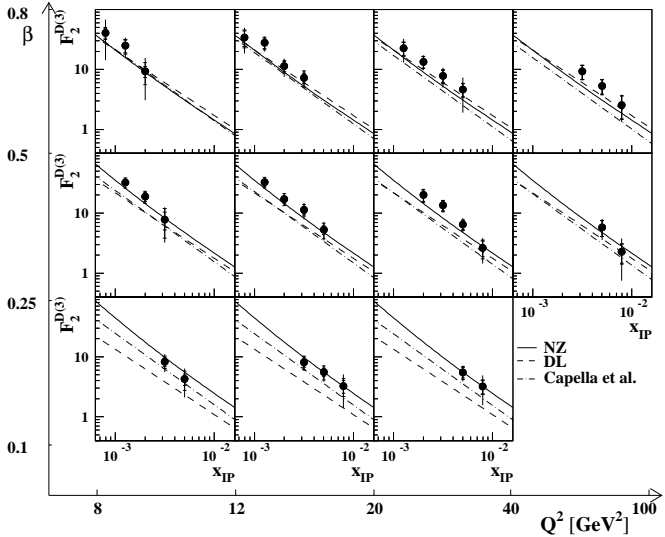}
\end{minipage} \ \
\begin{minipage}[b]{70mm}
\caption{\label{models}{\protect \sl
    $F_2^{D(3)}$ compared to various models discussed in the text. The
    estimated 15\% contribution due to double dissociation has been
    subtracted.  The overall normalization uncertainty of 3.5\% due to
    the luminosity and 10\% due to the subtraction of the double
    dissociation background is not included in th error bars.  }}
\vspace{15mm}
\end{minipage}
\end{figure}
The comparison with the various models discussed above presented in
figure ~\ref{models} shows that the models of Nikolaev-Zakharov and of
Capella et al. give the best description of the data. The model of
Donnachie--Landshoff reproduces the data well in the region of $\beta
> 0.25$ where the contribution of the $f$ trajectory is
negligible. The original model of Ingelman--Schlein (not shown), which
assumes that quarks saturate the momentum sum rule for the $\pom$
overestimates the cross section by a large factor.

It should be noted that the $F_2^{D(3)}$ dependence on $x_{\pom}$,
$\beta$ and $Q^2$ is also well reproduced by the model of Buchmueller
and Hebecker~\cite{Buchmueller} which does not use the concept of the
Pomeron, but rather describes the production of LRG through one gluon
exchange with a non-perturbative color cancellation induced by wee
partons.

\section{Diffractive jet production in hard photoproduction}

\begin{figure}[htb]
\begin{minipage}[b]{80mm}
\epsfxsize = 80mm
\epsfclipon
\epsffile[50 200 520 630]{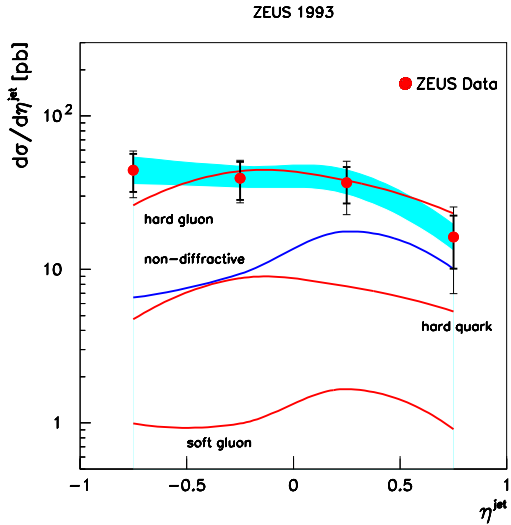}
\end{minipage} \ \
\begin{minipage}[b]{80mm}
\caption{\label{etajet}{\sl Measured differential $ep$ cross section
  $d\sigma/d\eta^{jet}(\etamax<1.8)$ for inclusive jet
  production.  The measurements are not corrected for the
  contributions from non-diffractive processes and double
  dissociation. The inner error bars represent the statistical errors
  of the data, and the total error bars show the statistical and
  systematic errors added in quadrature. The shaded band displays the
  uncertainty due to the energy scale of the jets, not included in the error
  bars.}}
\vspace{15mm}
\end{minipage}
\end{figure}

The first observation of the partonic substructure of diffractive
proton dissociation was reported by the UA8 experiment~\cite{Schlein}
in two jet production associated with a tagged leading proton.  An
analogous measurement was performed by ZEUS~\cite{ZEUSLRGPH, jetLRG2}
where large $p_T$ jet production was studied for photoproduction
events with a LRG, $\etamax<1.8$.  The inclusive cross section
$d\,\sigma/d\,\eta^{jet}$ was determined in the energy
range $135 \lsim W \lsim 280 \gev$ for $-1<\eta^{jet}<1$,
requiring the transverse jet energy $E_T>8 \gev$. The cross section
was then compared with expectations of a model in which jet production
with a LRG in the forward region is due to a hard scattering between
the partons of the pomeron and the photon. Both the direct photon and
resolved photon contribution were taken into account, with the
$\gamma$ structure function as given by the GS--HO
parameterization~\cite{gs}. The flux of the $\pom$ was assumed to be
given by the Donnachie--Landshoff parameterization~\cite{DLflux} and
the pomeron was assumed to consists of either only quarks or only
gluons, soft or hard, in each case saturating the momentum sum rule.
The results of the comparison are shown in figure~\ref{etajet}.
Also shown is the contribution expected from fluctuations of the
hadronic final state in non-diffractive jet photoproduction. The
excess of events with LRG is best described by a pomeron consisting of
hard gluons. It is clear though that a combination of gluons and
quarks would also reproduce the cross section.  The photoproduction
data alone cannot be used to decompose the content of the pomeron
without further assumptions about the flux of the $\pom$ and the
associated momentum sum rule.

\section{Partonic Content of the Pomeron}

\begin{figure}[htb]
\begin{minipage}[b]{80mm}
\epsfxsize = 80mm
\epsfclipon
\epsffile[60 200 490 620]{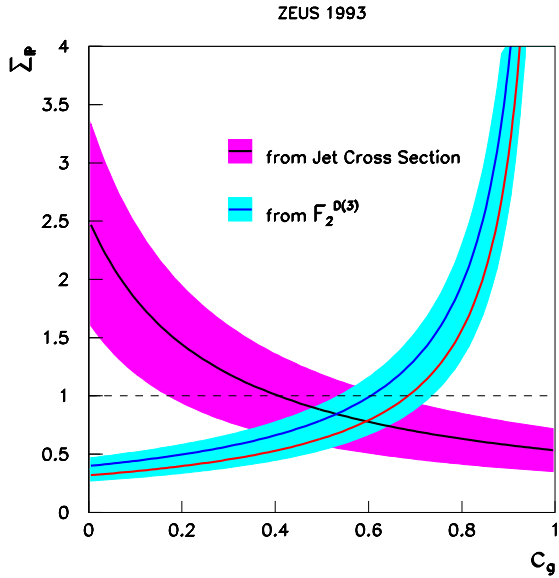}
\end{minipage} \ \
\begin{minipage}[b]{80mm}
\caption{\label{pomcon}{\sl
    The overall normalization coefficient $\Sigma_{\pom}$ required to
    describe the jet cross section and the $F_2^{D(3)}$ measurement,
    assuming a factorizable pomeron consisting of hard gluons carrying
    a $c_g$ fraction of its momentum and of quarks carrying a $1-c_g$
    fraction of its momentum.  }} \vspace{20mm}
\end{minipage}
\end{figure}

The DIS data on the $F_2^{D(3)}$ structure function, sensitive to the
quark content, and the jet photoproduction cross section, sensitive to
both the quark and gluon content, can be combined to estimate the
percentage of the $\pom$ momentum carried by the quarks and the
gluons. This is done in the following way. The pomeron flux is assumed
to be the same in photoproduction and in DIS and given by the
Donnachie--Landshoff parameterization. The shape of the quark
distribution is taken as determined from the DIS measurement and is
normalized to one. The gluon distribution is assumed to be of the form
$\beta(1-\beta)$ and also normalized to one. Then both the DIS and the
jet photoproduction cross sections are calculated assuming that the
gluons carry $c_g$ fraction of the pomeron momentum, the rest being
carried by the quarks. For each $c_g$ the overall normalization
$\Sigma_{\pom}$ needed to reproduce the measured cross section is
determined separately for the two reactions. The results are presented
in figure~\ref{pomcon}.
The region where the $\Sigma_{\pom}$ is the
same for both reactions determines the gluon content of the pomeron.
This happens for $0.3<c_g<0.8$, where the range is determined by the
statistical and systematic uncertainties on the cross section
measurements. Note that this result does not depend on the assumed
pomeron flux, nor on the contribution of the proton dissociation. It
is also independent of any assumption on the validity of the momentum
sum rule for the pomeron. The conclusion is thus that the gluon
content of the pomeron is substantial.

The same conclusion can be drawn from the lack of $Q^2$ evolution of
$F_2^{\pom}$ at large $\beta$, assuming that the DGLAP evolution
equation applies to the pomeron~\cite{Stirling}. Since, as is the case
for the proton, parton radiation leads to the depletion of quarks at
large $\beta$ ($x$ in the proton case), the only way to compensate
this depletion is by adding quarks from another source - gluon
radiation of quarks. These gluons have to be then at $\beta
\rightarrow 1$. The numerical estimate of this effect can be found
in~\cite{h1glu}.

\section{Vector meson production in DIS}

Another example of a reaction which leads to LRG in the final states
of electroproduction is the exclusive production of vector mesons~(see
contribution by J. Whitmore to this conference) such as $ep
\rightarrow e\rho p$.  The previously existing data from
photoproduction~\cite{rhogp} and electroproduction~\cite{rhoep} were
compatible with expectations based on the Pomeron phenomenology. In a
model proposed by Donnachie and Landshoff~\cite{rhodl} the $\pom$
consists of two non--perturbative gluons and the large $Q^2$ $\rho$
production cross section is predicted to have the following form:
\[
\sigma(\gamma^*p\rightarrow \rho^o p) \sim
\frac{\left(xg(x,Q^2)\right)^2}{Q^6} \, ,
\]
where the $x$ dependence of the $xg(x,Q^2)$ is determined by the
intercept of the soft $\pom$.  Thus the $\rho^o$ production cross
section is expected to rise only slowly with $W$,
$\sigma(\gamma^*p\rightarrow \rho^o p) \sim W^{0.32}$.
\begin{figure}[htb]
\begin{minipage}[b]{85mm}
\epsfxsize = 85mm
\epsfclipon
\epsffile[40 180 520 640]{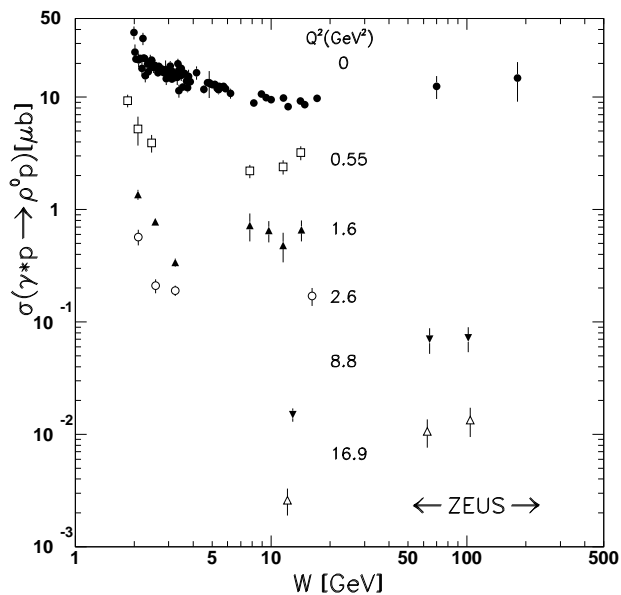}
\end{minipage} \ \
\begin{minipage}[b]{80mm}
\caption{\label{sigmarho}{\sl
    The $\gamma^* p \rightarrow \rho^o p$ cross section as a function
    of $W$, the $\gamma^*p$ center of mass energy, for several values
    of $Q^2$.  The ZEUS data at $Q^2$ = 8.8 and 16.9 GeV$^2$ have an
    additional 31\% systematic normalization uncertainty (not shown).
    }}
\vspace{20mm}
\end{minipage}
\end{figure}
This expectation is to be contrasted with the recent calculation of
Brodsky et al.~\cite{brodskyetal} which leads within perturbative QCD
to a similar formula as the one of Donnachie and Landshoff but where,
for a longitudinally polarized virtual photon at $t=0$, the gluon
distribution entering the formula is that of the proton, as measured
in DIS. In this case one expects the cross section to rise almost
linearly with $W$.

The ZEUS measurements of $\sigma(\gamma^*p\rightarrow \rho^o p)$ for
photoproduction~\cite{rhogpzeus} and for $7<Q^2<25 \gev^2$ and
$40<W<130 \gev$~\cite{rhoepzeus} are presented in
figure~\ref{sigmarho} and compared with measurements at lower
$W$. While the photoproduction cross section is slowly rising with
$W$, that for $Q^2>7 \gev^2$ exhibits a much stronger increase
relative to lower $W$ data.  In accordance with expectations the
contribution of longitudinally polarized photons was found to be
larger than that of the transversely polarized one (assuming $s$
channel helicity conservation). It is interesting to note that the
$Q^2$ dependence turns out to be $Q^{-a}$ with $a=4.2 \pm 0.8
^{+1.4}_{-0.5}$, again closer to the expectation of the perturbative
calculations in which the increase of the gluon density with $Q^2$
compensates partly the $Q^{-6}$ dependence due to the two gluon
exchange~\cite{halfms}.

\section{Conclusions}

The production of large rapidity gap events in $ep$ interactions at
HERA has opened a unique opportunity to study diffractive dissociation
in reactions associated with at least one hard scale, like the deep
inelastic scattering or production of large transverse momenta
jets. The emerging picture, if interpreted in terms of a factorizable
pomeron exchange, leads to a pomeron consisting of quarks and gluons
with a rather hard momentum distribution. The appearance of jets in
the deep inelastic scattering and the large cross section for vector
meson production measured at large $Q^2$ and large center of mass
energies could well be due to a perturbative mechanism, implying a
more complicated nature of the pomeron.

\end{document}